\documentclass[%
 reprint,
 amsmath,amssymb,
 aps,
pra,
]{revtex4-2}

\usepackage{graphicx}% Include figure files
\usepackage{dcolumn}% Align table columns on decimal point
\usepackage{bm}% bold math
\usepackage{blindtext}
\usepackage{graphicx}
\usepackage{braket}
\usepackage{color}
\usepackage{siunitx}
\usepackage{amsmath}
\usepackage{statmath}
\usepackage{bm, xspace}

\newcommand{\fm}[1]{\ifmmode#1\else$#1$\fi}

\newcommand{\HCI}[2]{#1\fm{^{#2+}}\xspace}
\newcommand{\Ca}[0]{\HCI{Ca}{14}}
\newcommand{\Be}[0]{\HCI{Be}{}}
\newcommand{\threePone}[0]{\fm{^3\mathrm{P}_1}\xspace}
\newcommand{\threePzero}[0]{\fm{^3\mathrm{P}_0}\xspace}
\newcommand{\vect}[1]{\boldsymbol{#1}\xspace}
\newcommand{\Bdc}{\vect{B_\mathrm{dc}}}
\newcommand{\Brf}{\vect{B_\mathrm{rf}}}

\begin{document}
	\title{Characterization of rf field-induced a.c.\ Zeeman shift in multi-level highly charged ions}
    
    \author{Shuying Chen$^1$} 
    \thanks{These authors contributed equally.}
    %\email{shuying.chen@quantummetrology.de}
   \author{Lukas J. Spieß$^1$}
    \thanks{These authors contributed equally.}
	\author{Alexander Wilzewski $^1$}
	\author{Malte Wehrheim$^1$}
	  \author{José R. Crespo López-Urrutia$^2$}
	\author{Piet O. Schmidt$^{1,3}$}
    \email{piet.schmidt@quantummetrology.de}
    
    \affiliation{$^1$Physikalisch-Technische Bundesanstalt, Braunschweig, Germany}
    \affiliation{$^2$Max-Planck-Institut für Kernphysik, Heidelberg, Germany}
	\affiliation{$^3$Institut für Quantenoptik, Leibniz Universität Hannover, Germany}

	\begin{abstract}
		%Highly charged ion clocks are an important optical clock platform for both frequency and timing metrology as well as fundamental physics tests.
        %Characterization of the trap rf induced a.c.\ Zeeman shift is essential for achieving high accuracy and stability of HCI-based optical clocks.        
        %In this work, we demonstrate the experimental characterization of the a.c.\ Zeeman shift of highly charged $\mathrm{Ca}^{14+}$. The transverse component of the a.c.\ magnetic field is measured by a modulation of the spectroscopy signal caused by the rf-induced a.c.\ magnetic field coupling the equally-spaced \threePone Zeeman manifold, which is then read out with quantum logic spectroscopy on the co-trapped $\mathrm{Be}^{+}$. The longitudinal component is measured from probing the \Be magnetic field-insensitive hyperfine transition $|F=2,m_F=0 \rangle \rightarrow | F=1,m_F=0 \rangle$.
        %We confirm the small influence of the a.c.\ Zeeman shift in highly charged ions and show that the employed techniques can easily be transferred to other multi-level atomic systems.
        Characterization of the trap rf induced a.c.\ Zeeman shift is essential for achieving high accuracy in optical ion clocks.        
        In this work, we demonstrate the experimental characterization of this shift using highly charged $\mathrm{Ca}^{14+}$. The transverse component of the a.c.\ magnetic field is measured using the Autler-Townes splitting of the equally-spaced Zeeman components of the \threePone when the Zeeman splitting is close to resonance with the trap rf drive frequency. We observe the resulting modulation by performing quantum logic spectroscopy using the co-trapped $\mathrm{Be}^{+}$. The longitudinal component is measured from probing the \Be magnetic field-insensitive hyperfine transition $|F=2,m_F=0 \rangle \rightarrow | F=1,m_F=0 \rangle$.
        We confirm the small influence of the a.c.\ Zeeman shift in highly charged ions. The employed techniques can easily be transferred to other multi-level atomic systems.
	\end{abstract} 
	
	\maketitle

    \section{\label{sec:I}Introduction}
    %--highly charged ion 
    Highly charged ions (HCIs) exhibit extreme electronic properties arising from their strong internal electric fields, which strongly suppress their sensitivity to external perturbations and make them promising systems for precision tests of fundamental physics \cite{safronova_search_2018, kozlov_hci}. In particular, several optical transitions in HCI have been predicted to possess among the highest sensitivities to variations of the fine-structure constant \cite{berengut_enhanced_2010, dzuba_highly_2015}, and to violations of local Lorentz invariance \cite{shaniv_new_2018, safronova_highly_2014}. 
    Moreover, precise measurements of atomic parameters in few-electron HCIs are of special interest, as theoretical calculations for such systems can achieve accuracies far beyond those attainable in many-electron atoms \cite{gilles_quadratic_2024, cheung_finding_2025}. 
    These properties make HCIs also a compelling platform for optical clocks, with applications in frequency and timing metrology as well as tests of fundamental physics \cite{king_optical_2022, wilzewski_nonlinear_2024}. Experimentally, a HCI-based optical clock has become possible by confining and sympathetically cooling of individual HCI in a linear Paul trap, allowing for quantum logic readout \cite{micke_coherent_2020, schmidt_spectroscopy_2005, schmoger_coulomb_2015}. In particular, HCI clocks are predicted to exhibit exceptionally small second-order Zeeman shifts \cite{gilles_quadratic_2024, dzuba_highly_2015}. This prediction has been experimentally confirmed by measurement of the quadratic Zeeman coefficients in Ref.~ \cite{g_factor}. A full evaluation of this shift requires knowledge of the d.c. and a.c. components of the magnetic field. The d.c. field can be obtained from measuring different Zeeman components of the clock transition. In ion traps, the dominant contribution to the a.c. field stems from the rf drive of the Paul trap, which induces currents at the trap drive frequency. %$\Omega_\mathrm{rf}$, 
    %$\vect{B_{\mathrm{rf}}}$. 
    This a.c.\ magnetic field gives rise to second-order Zeeman shifts in optical clocks \cite{rosenband_frequency_2008, gan_oscillating-magnetic-field_2018} and perturbs precision measurements of Landé $g$-factors \cite{arnold_precision_2020, rehmert_lande_2025}.
    The trap-drive-induced a.c.\ magnetic field can be decomposed into longitudinal and transversal components.
    % of $B_\mathrm{rf}$. %three orthogonal components with respect to the quantization axis defined by the static magnetic field: %$\vect{B_{\mathrm{dc}}}$, 
    %one component parallel to the static magnetic field and two perpendicular components. 
    %The parallel component couples the different fine levels or hyperfine levels with the same magnetic quantum number, which is first-order magnetic field insensitive. Therefore, it can be measured via measuring the fine or hyper-fine transition frequency at different rf power.
    %
    %The parallel component mostly couples states belonging to the different fine or hyperfine manifold with the same magnetic quantum number, whereas the perpendicular fields couple the different Zeeman sublevels within one fine or hyperfine manifold.
    %A standard technique has been to vary the rf confinement and extrapolate any measurable difference to zero. 
    In Ref.~\cite{Gan2018}, Autler-Townes (AT) splitting spectroscopy was introduced as a high-precision method to measure the transverse component of the a.c.\ magnetic field.
    Since then, this technique has been successfully applied to singly charged ion systems including \HCI{Yb}{} \cite{tofful_171yb+_2024}, \HCI{Ba}{} \cite{arnold_precision_2020}, and \HCI{Ca}{} \cite{ma_precision_2024, zhang_liquid_2025}, \HCI{Sr}{} \cite{lindvall_88Sr+_2025}. 
    %
    %However, the a.c.\ magnetic field in highly charged ion (HCI) systems has not yet been experimentally measured. 
    %All previous demonstrations have observed the AT splitting when only two levels are coupled by the field. 
    To generate a two-level system, either a state with only two Zeeman sublevels is chosen, e.g. $\mathrm{S}_{1/2}$, or a non-negligible second-order Zeeman shift yields an unequal splitting of neighboring Zeeman sublevels.
    While the former is only possible  for certain selected ions, the later condition is generally not satisfied for levels in HCI due to their much smaller second-order magnetic-field sensitivity coefficients \cite{g_factor}. As a result, the Zeeman sublevels within a given fine-structure level (for total angular momentum $J>0$) are usually nearly equally spaced at the relevant static magnetic fields, leading to a multi-level interacting system rather than an effective two-level system. This is also true for some singly charged ion species with low second-order magnetic field sensitivity, such as \HCI{B}{}, \HCI{Al}{}, and \HCI{In}{}.

\begin{figure*}[!htbp]
		\centering
        \includegraphics[width=\textwidth]{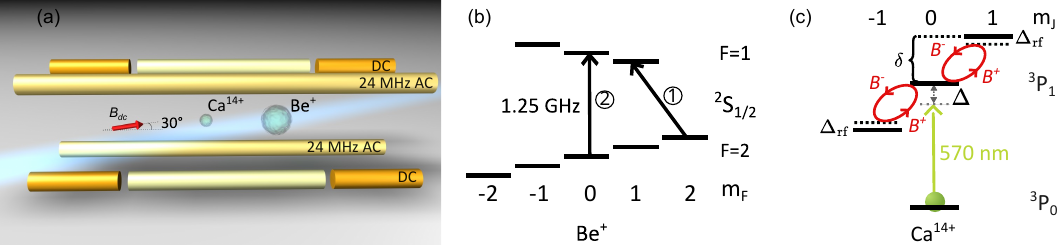}
		\caption{
        (a) A \Be-\Ca\ two-ion crystal confined in a linear Paul trap. The quantization axis at an angle of ca. $30^\circ$ relative to the symmetry axis of the trap is defined by $\Bdc$, which is also the direction of $B_z$. %The propagation of the Doppler-cooling laser is parallel to $\Bdc$. Coordinates $(x',y',z')$ correspond to the trap axial axis direction and two radial direction, and coordinates $(x,y,z)$ denote the direction of the magnetic field.  
        (b) Relevant energy levels in \Be for microwave manipulation at 1.25 GHz. Measurement of transition \textcircled{1} from edge state $|F=2, \mathrm{m}_{F} = 2\rangle \rightarrow |F=1, \mathrm{m}_{F}=1\rangle$ is used for magnetic field calibration. A measurement of the transition \textcircled{2} $|F=2, \mathrm{m}_{F} = 0\rangle \rightarrow |F=1, \mathrm{m}_{F}=0\rangle$ is used to determine $B_z^2$. (c) Relevant energy levels of \Ca involved in coupling to $B_\mathrm{+}$ ($B_\mathrm{-}$) and read out by the probe beam. $\Delta_{\mathrm{rf}}$ is the detuning of the a.c.\ magnetic field frequency $\Omega_{\mathrm{rf}}$ from the Zeeman splitting $\delta$: $\Delta_{\mathrm{rf}}=\delta-\Omega_{\mathrm{rf}}$. $\Delta$ is the detuning of the probe laser frequency to transition $\ket{\threePzero, m_J=0 } \rightarrow \ket{\threePone, m_J=0}$.
        }
		\label{fig:setup}
	\end{figure*} 
    
    % summary of our work  
    In this work, we demonstrate a measurement of the trap-induced a.c.\ magnetic field using highly charged \Ca and co-trapped \Be. The measurement principle is introduced in Section~\ref{sec:II}.
    In Section~\ref{sec:III}, we determine the transversal components of the a.c.\ magnetic field via spectroscopy of the resonant coupling among the three Zeeman sub-levels of the \threePone state in \Ca. 
    %Unlike previous such measurements in singly charged ions \cite{gan_oscillating-magnetic-field_2018}, the second-order Zeeman shift in \Ca can be neglected \cite{g_factor} and we observe resonant interaction between all three Zeeman substates in \threePone. 
    The coupling strengths are extracted by fitting the observed spectra to a three-level interaction model, from which the amplitudes of the transversal a.c.\ magnetic fields are derived.
    % The other orthogonal perpendicular magnetic component is also measured by flipping the dc magnetic field direction.
    %
    In Section~\ref{sec:IV}, the longitudinal component of the a.c.\ magnetic field is inferred by measuring the second-order Zeeman shift of the first-order magnetic-field-insensitive hyperfine transition $|F=2,m_F=0 \rangle \rightarrow |F=1,m_F=0\rangle$ in \Be.
    In Section~\ref{sec:V}, we estimate the ac-magnetic-field-induced second-order Zeeman shift under our typical HCI clock operating conditions and summarize its impact on clock accuracy. 
    
% summary of our work  
    %In this paper, we measure $\vect{B_{\mathrm{+}}}$ and $\vect{B_{\mathrm{-}}}$ through the energy level shifts from resonant coupling of Zeeman sublevels of the \threePone state in highly charged \Ca. Unlike previous such measurements in singly charged ions \cite{gan_oscillating-magnetic-field_2018}, the second-order Zeeman shift in \Ca can be neglected \cite{g_factor} and we observe resonant interaction between all three Zeeman substates in \threePone. The last component, $\vect{B_{z}}$ is derived from measuring the second-order Zeeman shift in \Be and its known second-order Zeeman effect coefficient \cite{g_factor}.

\section{\label{sec:II} Measurement Principle}
The experiments here are performed on a \Be-\Ca two-ion crystal prepared and confined in a cryogenic linear Paul trap \cite{leopold_cryogenic_2019, micke_closed-cycle_2019,g_factor, king_optical_2022, wilzewski_nonlinear_2024} as sketched in Fig.~\ref{fig:setup}(a). \Ca is the spectroscopy ion and \Be is the logic ion used for sympathetic cooling to the motional ground state and quantum logic readout of the electronic state of \Ca \cite{micke_coherent_2020, schmidt_quantum_2009}. The relevant hyperfine transitions in \Be $\ket{\mathrm{S}_{1/2}, F=2} \rightarrow \ket{\mathrm{S}_{1/2}, F=1}$, specifically transition \textcircled{1} and transition \textcircled{2}, are shown in Fig.~\ref{fig:setup}(b). They have a well-known magnetic response \cite{wineland_laser-fluorescence_1983, shiga_diamagnetic_2011} and are driven by a microwave emitted by an antenna close to the ion trap at \SI{1.25}{\giga\hertz}. The relevant optical transition $\threePzero \rightarrow \threePone$ in \Ca is shown in Fig. \ref{fig:setup}(c), and is driven by a Hz-level ECDL at \SI{570}{\nano\meter} \cite{wilzewski_nonlinear_2024}. This laser is pre-stabilised to a high-finesse cavity \cite{webster_force-insensitive_2011, dawel_coherent_2024}, then further stabilized \cite{stenger_ultraprecise_2002} to the ultrastable 'Si2' cavity at PTB \cite{matei_1.5_2017} through an optical frequency comb. 
%The relevant energy levels for \Be and \HCI{Ca}{14} are shown in Fig.~\ref{fig:setup}(b) and (c). 
%The experimental setup has been described in previous publications \cite{g_factor, king_optical_2022, wilzewski_nonlinear_2024} and will only be summarized here. 
%
The rf trap drive $\Omega_{\mathrm{rf}}/(2\pi)$ of the linear Paul trap is set at about \SI{24.2}{\mega\hertz} and provides radial confinement. It also generates an a.c.\ magnetic field $\vect{B_{\mathrm{rf}}}$.

Three orthogonal pairs of coils set the magnetic field direction $\vect{B_{\mathrm{dc}}}$ and define the quantization axis at an angle of approximately \SI{30}{\degree} relative to the symmetry axis of the trap, as shown in Fig.~\ref{fig:setup}(a). The field strength can be enhanced by optionally adding NdFeB permanent magnets outside the vacuum chamber.
The static magnetic field $\vect{B_{\mathrm{dc}}}$ is measured using a commercial fluxgate magnetometer outside the vacuum chamber. An additional set of compensation coils is used to actively stabilize the measured magnetic field to a relative stability of $10^{-5}$ for up to 100~s at the ion position \cite{leopold_cryogenic_2019}.

The trap-drive-induced a.c.\ magnetic field $\Brf$ can be decomposed into three orthogonal components in a cylindrical basis with respect to the $z$-axis defined by the static magnetic field $\vect{B_{\mathrm{dc}}}$: 
    \begin{equation}
        \vect{B_{\mathrm{rf}}} =  (B_{+}, B_-, B_z), \label{eq:BRF}
    \end{equation}
    
%The components $\vect{B_{x}}$ and $\vect{B_{y}}$ are from the components transverse to the quantization axis, 
where we have defined the spherical components $B_{\mathrm{+}}$ and $B_{\mathrm{-}}$. The remaining component $B_{z}$ is parallel to the quantization axis. 
%The orientations of $\vect{{x,y,z}}$ axes are illustrated in Fig.~$\ref{fig:setup}$. 

%The static magnetic field $\vect{B_{\mathrm{dc
%}}}$ is measured via the first-order magnetic field sensitive transition $| F=2, mF=2 \rangle \rightarrow | F=1, mF=1 \rangle$ in \Be, as shown in Fig.~\ref{fig:setup}(b). The amplitude of $\vect{B_{\mathrm{dc
%}}}$ is then derived from Breit-Rabi formula together with the well-known $g$-factor \cite{shiga_diamagnetic_2011}. 

%Since the a.c.\ magnetic field depends on the trap drive power $P_\mathrm{rf}\propto B_{\mathrm{rf}}^2$, we normalize $B_\mathrm{rf}$ to the radial motional frequency $\omega_r$ of a single \Be\ using $P_\mathrm{rf} \propto \omega_r^2$. (put at somewhere else) 

Similar to the method described in Ref.~\cite{Gan2018}, the amplitude of the transverse magnetic-field components $B_{\mathrm{+}}$ and ${B_{\mathrm{-}}}$ can be measured by observing the AT splitting from resonant coupling between neighboring Zeeman states within a fine- or hyperfine-structure manifold when the Zeeman splitting $\delta=g\mu_B B_{dc}$ approaches the trap drive frequency $\Omega_\mathrm{rf}$, as shown in Fig.~\ref{fig:setup}(c). Here, $g$ is the $g$-factor and $\mu_B$ is the Bohr magneton.
%For the co-trapped \Be, this resonance occurs at $B_{dc}=$\SI{3.6}{\milli\tesla}, which exceeds the maximum magnetic field achievable in our system. 
We choose the excited state \threePone\ of \Ca, shown in Fig.~\ref{fig:setup}(c), since the resonance condition is satisfied at $B_{dc}=$\SI{1.15}{\milli\tesla}, requiring a smaller field when compared to \Be for which the resonance occurs at $B_{dc}=$\SI{3.6}{\milli\tesla}. %The coupling of the transverse components $B_{\pm}$ to the Zeeman sublevels of the \threePone state is depicted in Fig.~\ref{fig:setup}(c). 
The second-order Zeeman shift induced by $\vect{B_{\mathrm{dc}}}$ is on the order of a few hertz \cite{g_factor} and is therefore negligible compared to the Rabi frequency associated with the a.c.\ magnetic field. Under these conditions, the \threePone\ manifold can be treated as a three-level system driven by $\vect{B_{\mathrm{rf}}}$. 
To this end, the DC coils in combination with NdFeB permanent magnets are used to generate a magnetic field $\vect{B_{\mathrm{dc}}}$ of up to $\pm$\SI{1.17}{\milli\tesla}. 
Here, the sign of $\vect{B_{\mathrm{dc}}}$ denotes its orientation, either aligned with $(+)$ or opposite to $(-)$ the Doppler-cooling laser propagation direction.
To measure the $B_{+}$ and $B_{-}$ separately, $\vect{B_{\mathrm{dc}}}$ is first aligned along the Doppler cooling laser propagation direction to determine $B_{+}$, as illustrated in Fig.~\ref{fig:setup}~(a). 
The field direction of $\vect{B_{\mathrm{dc}}}$ is then reversed, and the same measurement procedure is applied to extract $B_{-}$.

%the d.c. second-order Zeeman shift on resonance is Hz-level \cite{g_factor} which can be neglected compared to the Rabi frequency of the a.c.\ magnetic field. 
%As shown in Fig.~\ref{fig:setup}(c), when the Zeeman splitting $\delta$ approaches the frequency of $B_{\mathrm{rf}}$, $\delta/2\pi \approx \Omega_{\mathrm{rf}}/2\pi \approx  \SI{24.2}{\mega\hertz}$, the interaction with $B_{\mathrm{rf}}$ induces coherent coupling between the three Zeeman sublevels, resulting in an energy-level splitting \cite{Gan2018}. 
%

The excitation spectrum of the $|$\threePone $m_F = 0 \rangle$ state is then probed via the optical transition $|$\threePzero $m_F=0 \rangle$ $\rightarrow$ $|$\threePone\ $m_F=0 \rangle$ using quantum logic spectroscopy \cite{micke_coherent_2020, king_optical_2022}.

The coupling of \Ca levels to the $B_\mathrm{+}$ field and to the probe light field is described by the interaction Hamiltonian,
\begin{equation}
\begin{split}
    H = &\;\Omega_0 \hat{\sigma}_{+}^{(0)} e^{-i\Delta t_p} 
      +  \Omega_{\text{+}} \hat{\sigma}_{+}^{(0\rightarrow 1)} e^{-i(\Delta + \Delta_\mathrm{rf} )t_p} \\
      &+ \Omega_{\text{+}}  \hat{\sigma}_{+}^{(-1\rightarrow 0)}e^{-i(\Delta - \Delta_\mathrm{rf} )t_p}+ h.c..
     \end{split}
     \label{eq:H}
\end{equation}
Here $\Omega_0$ is the on-resonance Rabi frequency of the probe beam at \SI{570}{\nano\meter} coupling $|^{3}\mathrm{P}_0, m_J=0 \rangle \rightarrow |^{3}\mathrm{P}_1, m_J=0 \rangle$ and $\Delta$ its detuning from the unperturbed resonance, while $t_\mathrm{p}$ is the probe time. 
%$\Delta$ is the detuning of the probe frequency to the transition $|$\threePzero, $m_J=0 \rangle \rightarrow |$ \threePone, $m_J=0 \rangle$ resonance frequency.
$\hat{\sigma}_{+}^{(0)}$ is the spin excitation operator for the transition $|^{3}\mathrm{P}_0, m_J=0 \rangle \rightarrow |^{3}\mathrm{P}_1, m_J=0 \rangle$. $\hat{\sigma}_{+}^{(0\rightarrow 1)}$ and $\hat{\sigma}_{+}^{(-1\rightarrow 0)}$ correspond to the excitation operator within the \threePone sublevels: $|^{3}\mathrm{P}_1, m_J=0 \rangle \rightarrow |^{3}\mathrm{P}_1, m_J=1 \rangle$ and  $|^{3}\mathrm{P}_1, m_J=-1\rangle \rightarrow |^{3}\mathrm{P}_1, m_J=0 \rangle$, respectively. $\Omega_+$ is the on-resonant Rabi frequency of $B_+$ coupling the transition $|^{3}\mathrm{P}_1, m_J=0 \rangle \rightarrow |^{3}\mathrm{P}_1, m_J=\pm 1 \rangle$.
$\Delta_{\mathrm{rf}}=\delta-\Omega_\mathrm{rf}$ is the detuning of the a.c. magnetic field frequency from the Zeeman splitting, as shown in Fig.~\ref{fig:setup}(c).
The quantum logic readout is not included here.

The ion is initially prepared in the ground state \threePzero.
The time evolution of the involved electronic-state populations in \Ca is described by the Liouville equation:
\begin{equation}
    \Dot{\rho}(t) = -\frac{i}{\hbar}[H,\rho(t)] .
    \label{eq:l_eq}
\end{equation}
The experimental observable is the excitation probability $1-\rho_0$, where $\rho_0$ denotes the population remaining in the ground state \threePzero. 
Equation~(\ref{eq:l_eq}) is solved numerically using the QuTiP package \cite{johansson_qutip_2013}, and fitted to the measured spectra to extract the coupling strength, from which the amplitude $B_+$ is determined. 
The analysis for the $B_-$ component follows an analogous procedure.

\section{\label{sec:III} Measurement of ${B_{\mathrm{+/-}}}$}     	
\begin{figure}[htpb]
		\centering	\includegraphics[width=\columnwidth]{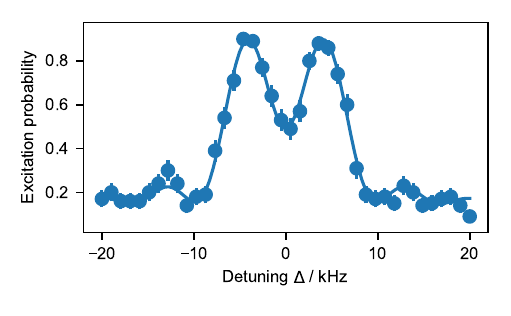}
		\caption{Typical spectrum of the $m_J=0\rightarrow m_{J'}=0$ transition in \Ca\ near resonance of the trap-drive-induced a.c.\ Zeeman shift, i.e. $\Delta_\mathrm{rf}\approx 0$. The detuning is relative to the unperturbed transition frequency. The solid line is a fit to derive $B_+$ or $B_-$.}
		\label{fig:at_signal}
	\end{figure}
    
A typical excitation spectrum of the optical transition in \Ca near-resonance condition $\Delta_\mathrm{rf}\approx 0$ as a function of laser detuning $\Delta$ is shown in Fig.~\ref{fig:at_signal}. The observed splitting in the spectrum arises from the coupling induced by ${B_+}$. The solid line shows a fit to the data using Eqs.~(\ref{eq:H}) and (\ref{eq:l_eq}) with $\Omega_+$, the excitation amplitude and offset as free parameters. The fit shows good agreement with the experimental data. 

\begin{figure}[!htbp]
    \centering      
    \includegraphics[width=\columnwidth]{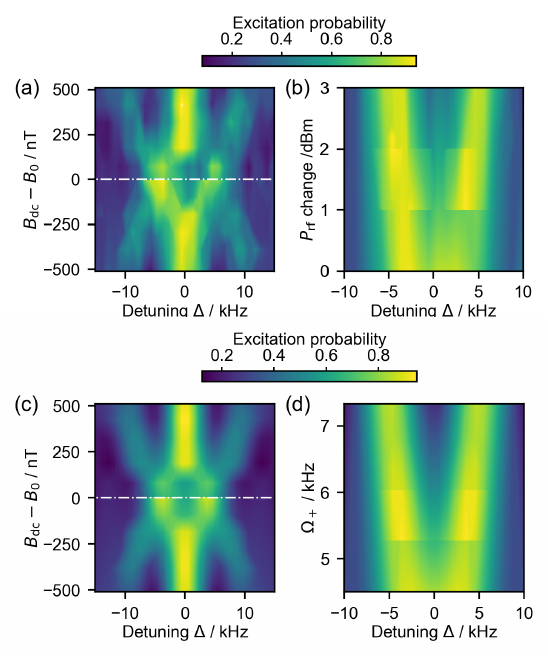}
    \caption{ (a) Experimental excitation probability for changing probe laser detuning $\Delta$ and static magnetic field $B_{\text{dc}}$. The white dashed line marks the resonant static magnetic field for $\Delta_\mathrm{rf} = 0$.  (b) Experimental excitation probability for changing probe laser detuning $\Delta$ and rf power $P_{\text{rf}}$. (c) and (d) show the predictions of the theory model using the fit parameters to the experimental data.}
    %Simulation for $\Omega_{\rm rf}=2\pi\times$\SI{5.5}{\kilo\hertz}, $\Omega_0=2\pi\times$\SI{4.16}{\kilo\hertz} and $t_p=$\SI{150}{\micro\second}, where the latter two values correspond to the experimentally measured ones.}
    \label{fig:MagnDependence}
\end{figure}

%When the magnitude of $B_{\mathrm{dc}}$ is varied while the rf power is fixed, the detuning between $\Omega_\mathrm{rf}$ and the Zeeman splitting $\Delta_{\text{rf}}$ changes linearly with $B_{\mathrm{dc}}$. 
Fig.~\ref{fig:MagnDependence}(a) shows the experimentally measured excitation spectra obtained by scanning the laser detuning $\Delta$ at different values of $\Bdc$, while keeping the trap-drive rf power $P_\mathrm{rf}$ fixed, as indicated on the vertical axis. %The magnetic field  $B_{\mathrm{dc}}$ is set by the changing the coils current and calibrated with the microwave transition in \Be. 
By fitting all spectra using Eqs.~(\ref{eq:H}) and (\ref{eq:l_eq}) with $\Delta_\mathrm{rf}$, excitation contrast and offset as free parameters, the corresponding values of $\Delta_\mathrm{rf}$ are extracted for each $\Bdc$.
%And from the relation $\hbar \Delta_\mathrm{rf}=\mu_B g (\Bdc-B_0)$, the static magnetic field $B_0$ corresponding to the resonance condition $\Delta_\mathrm{rf}=0$ is then determined to be $B_0=\SI{1153.18} {\nano\tesla}$ in our experiment, marked by the white dashed line in Fig.~\ref{fig:MagnDependence}(a). Here the $g-$factor for \threePone in \Ca is 1.49903, measured in Ref.~\cite{g_factor}. 
Fig.~\ref{fig:MagnDependence}(c) shows the spectra reconstructed from the fit parameters, demonstrating good agreement with the experimental spectrum in Fig.~\ref{fig:MagnDependence}(a).

%Setting the magnetic field $B_\mathrm{dc}$ to the resonance, we vary the rf power $P_\mathrm{rf}$. This changes the coupling strength characterized by the on-resonant Rabi frequency $\Omega_\mathrm{ac} \propto \sqrt{P_\mathrm{rf}}$. The signals for different $P_\mathrm{rf}$ are shown in Fig.~\ref{fig:MagnDependence}(b). 

With the magnetic field $\Bdc$ set to the resonance condition, white dashed line in Fig.~\ref{fig:MagnDependence}(a), we determine the dependence of $B_\mathrm{+}$ on the applied trap drive power $P_\mathrm{rf}$, characterized by the measured radial motional frequency $\omega_r\propto\sqrt{P_\mathrm{rf}}$, to determine the a.c.\ magnetic field in the rf power range of interest. 
%With the magnetic field $\Bdc$ set to the resonance condition, the rf power $P_\mathrm{rf}$ is varied. 
%This changes the on-resonant Rabi frequency $\Omega_\mathrm{+} \propto \sqrt{P_\mathrm{rf}}$. 
The spectra recorded for different values of $P_\mathrm{rf}$ are shown in Fig.~\ref{fig:MagnDependence}(b). 
%The yellow area on the graph shows the change in excitation peak splitting, which increases with increasing $P_\mathrm{rf}$.
By fitting the spectra using Eqs.~(\ref{eq:H}) and (\ref{eq:l_eq}) with $\Omega_\mathrm{+}$, excitation amplitude and offset as free parameters, the corresponding values of $\Omega_\mathrm{+}$ are extracted for each $P_\mathrm{rf}$.
Fig.~\ref{fig:MagnDependence}(d) shows the spectra reconstructed from the fit parameters, which are in good agreement with the experimental data. 
The magnitude of $B_\mathrm{+}$ is obtained from $\Omega_\mathrm{+}$ using \cite{gan_oscillating-magnetic-field_2018}
\begin{align}
    B_\mathrm{+} = \frac{\sqrt{2}\hbar \Omega_\mathrm{+}}{\mu_\mathrm{B} g}.
\end{align}

\begin{figure}[!htbp]
    \centering    
    \includegraphics{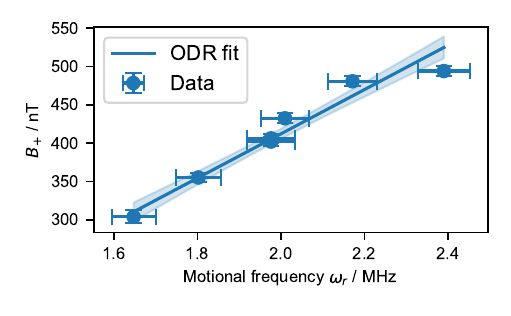}
    \caption{Extracted values of $B_{\mathrm{+}}$ as a function of the motional frequency $\omega_r$ at varying rf powers. The solid line represents the orthogonal distance regression fit, and the shaded region denotes the $1\sigma$ confidence interval.}
\label{fig:split_RF_power1}
\end{figure}

%The characterization of $P_\mathrm{rf}$ reaching the trap is done by relating it to the radial motional frequency $\omega_r$, which scales with $\sqrt{P_\mathrm{rf}}$. 
%The radial frequency is measured by coherently exciting the out of phase mode using an rf electric field from a spare electrode \cite{g_factor}. 
%
The relationship between $B_{\mathrm{+}}$ and the experimentally accessible $\omega_r$ is shown in Fig.~\ref{fig:split_RF_power1}. The solid blue line represents an orthogonal distance regression fit with a reduced $\chi^2$ of 0.96.  We define the secular-frequency–normalized magnetic field as $B_{\mathrm{+,norm}}= B_{\mathrm{+}}/\omega_r$. The extracted slope is $B_{\mathrm{+,norm}} = \SI{288\pm 31}{\nano\tesla\per\mega\hertz}$.
The light blue shaded area indicates the $1\sigma$ confidence interval. The same definition is used for $B_{\mathrm{-,norm}}$, $B_{\mathrm{z,norm}}$, $B_{\mathrm{rf,norm}}$ throughout the remainder of the text.
%The x axial error bar on the data points is from motional frequency measurement uncertainty and y axial error bar is too small to be seen. 
%shown as Fig.~\ref{fig:split_RF_power1}(b) 

Subsequently, the same measurement and analysis were also performed for $B_{\mathrm{-}}$ with the $\vect{B_{\mathrm{dc}}}$ direction reversed, yielding a fitted linear slope of $B_{\mathrm{-,norm}} = \SI{261\pm 95}{\nano\tesla\per\mega\hertz}$. The larger uncertainty is from a larger statistical error. %Nevertheless, it allows us to estimate the $B_{\mathrm{rf}}$-induced second-order Zeeman shift to be negligible.
%, making the absolute uncertainty not important. 
     
\section{\label{sec:IV} Measurement of ${B_{z}}$}
The parallel component ${B_{z}}$ is measured through the transition frequency of the hyperfine transition transition \textcircled{2} in single \Be\ as shown in Fig.~\ref{fig:setup}(b). This transition is only second-order magnetic field sensitive with a well-known coefficient \cite{wineland_laser-fluorescence_1983, shiga_diamagnetic_2011}, which can be written as \cite{gan_oscillating-magnetic-field_2018}
    \begin{equation}
        \delta\nu_{0\rightarrow0} \approx 3.141\times 10^{5}\left( B_{\mathrm{dc}}^2 + \langle B_z^2\rangle + \frac{\langle B_+^2\rangle}{4} + \frac{\langle B_-^2\rangle}{4}\right)~\frac{\mathrm{MHz}}{\mathrm{T}^2},
    \end{equation}
where $\langle \cdot \rangle$ indicates a temporal average. %
The first term in the bracket arises from the static magnetic field $B_{\mathrm{dc}}$ while the remaining terms are from the rf magnetic field $\vect{B_{\mathrm{rf}}}$, where the induced shift is proportional to $B_{\mathrm{+,-,z}}^2$, and therefore scales with $\omega_r^2$. 
To determine the static magnetic field $B_{\mathrm{dc}}$, we probe the first-order magnetic field sensitive transition \textcircled{1} in \Be, as shown in Fig.~\ref{fig:setup}(b). $B_{\mathrm{dc}}$ is then calculated from the measured transition frequency using its first-order magnetic field sensitivity \cite{rosenband_frequency_2008}. 
    
Experimentally, both microwave transitions are measured interleaved with a two-point sampling lock \cite{peik_laser_2006} using Rabi excitation. The probe durations for each measurement are \SI{92}{\milli\second} for transition \textcircled{2} and \SI{446}{\micro\second} for \textcircled{1}. The measured statistical uncertainty of the frequency of transition \textcircled{2} is approximately 10~mHz. 
Both transition frequencies are recorded for different $P_\mathrm{rf}$, characterized again through $\omega_r^2$.  
    
\begin{figure}[htpb]
		\centering
		\includegraphics[width=\columnwidth]{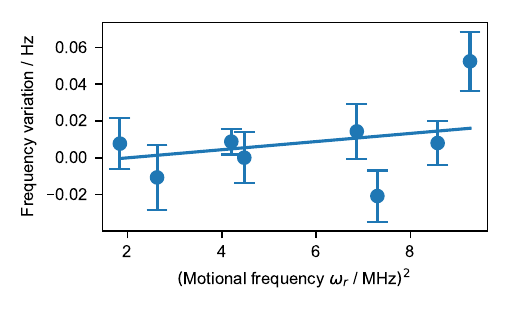}
		\caption{Relative transition frequency of transition \textcircled{2} in \Be\ as a function of $\omega_r^2$, which characterizes the rf power $P_\mathrm{rf}$ in the trap. The error bars represents the statistical uncertainties.}
		\label{fig:aczeemanpi}
\end{figure}

The change in $\delta\nu_{0\rightarrow0}$ as a function of $\omega_r^2$ is shown in Fig.~\ref{fig:aczeemanpi}. 
A linear fit yields
\begin{equation}
\begin{aligned}
    B_{\mathrm{rf, norm}} &= \sqrt{\langle (B_{\mathrm{z,norm}})^2\rangle + \frac{\langle (B_{\mathrm{+,norm}})^2\rangle}{4} + \frac{\langle (B_{\mathrm{-,norm}})^2\rangle}{4}}\\
    &= \SI{84\pm36}{\nano\tesla\per\mega\hertz},
\end{aligned}
\label{eq:Brf_norm}
\end{equation}
with a reduced $\chi^2$ of 2.05. Considering the above measured $B_{\mathrm{+,norm}}$ and $B_{\mathrm{-,norm}}$, this indicates that $B_{\mathrm{z,norm}}=$ \SI{80\pm49}{\nano\tesla\per\mega\hertz} in our trap, with the uncertainty derived using Monte-Carlo sampling (see Appendix~\ref{appendix} for details), and unphysical results $\langle B_{\mathrm{z,norm}}^2\rangle<0$ excluded.
    
%This value is similar to observation in other linear Paul traps \cite{gan_oscillating-magnetic-field_2018} \note{cite Johannes? other traps?} when rescaling the typically applied voltages. 

\section{\label{sec:V} Conclusion}
We demonstrate the measurement of the trap rf–induced magnetic field by observing the coherent interaction of the a.c.\ magnetic field with the \threePone state in \Ca as well as the second-order Zeeman shift in \Be. The former revealed a direct interaction between three Zeeman sublevels. The measurements yielded an a.c.\ magnetic-field amplitude of $\vect{B_{\mathrm{rf, norm}}}=~\left( 288(31),\, 261(95),\,80(49)\right)$~$\mathrm{nT~MHz^{-1}}$, where the first two terms correspond to the spherical components $B_{\mathrm{+, norm}}$ and $B_{\mathrm{-,norm}}$, and the third value represents the component $B_{\mathrm{z,norm}}$ parallel to the quantization axis.
%
%The measured a.c.\ magnetic field is smaller than that typically reported for optical clocks based on singly charged ions \cite{gan_oscillating-magnetic-field_2018, arnold_precision_2020, kramer_thesis_2022, brewer_measurements_2019}. This reduction arises partly from the large charge-to-mass ratio of the HCI, which enables ion trapping with lower voltages than those generally required for singly charged ions. 

Due to the large charge-to-mass ratio of the HCI, stable trapping can be achieved at lower trapping rf voltages than those required for singly charged ions. Consequently, a reduced a.c.\ magnetic field is observed under our typical operating conditions. 
When the trapping rf voltage is rescaled to that used for singly charged ions at comparable trapping frequencies, the inferred a.c.\ magnetic field in our trap is comparable with values reported in other optical clock systems \cite{gan_oscillating-magnetic-field_2018, arnold_precision_2020, kramer_thesis_2022, brewer_measurements_2019}. 
%Comparison to other literature:
%Arnold 2020: 0.8*micro*tesla*(1.6*24*9/(0.890*130*20.7)) = 0.1µT (their field * our sec. freq. * our rf freq. * our mass / (their sec. freq. *their mass* their rf freq) = our field , thus ours higher by factor of 4
%Gan 2018: 4.5*micro*tesla*(1.6*24*9/(0.5*1171*30)) = 0.6µT , larger than us by factor of 1.5
%Johannes Diss 2022: 27µT*(1.6*24*9/(1.5*40*28))=4µT, here smaller by factor of 20
%private communication Iqloc3 (rough first estimate): 11.3*micro*tesla*(1.6*24*9/(2.5*40*28))=1.3µT, larger by a factor of 4

For the clock transition \threePzero $\rightarrow$ \threePone in \Ca, the measured a.c.\ magnetic field results result in a fractional second-order Zeeman shift below $10^{-22}$ \cite{g_factor, gilles_quadratic_2024}, rendering it negligible for clock operation. This insensitivity benefits from the small second-order Zeeman coefficient characterized in \cite{g_factor}. 
%
%Similarly, for the excited state $g$-factor measurement in \Ca described in Ref. \cite{g_factor}, the contribution from the a.c.\ magnetic field is below $10^{-7}$ level, and therefore is also negligible.

These results demonstrate that, for optical clocks based on HCI, the requirements on the characterization of the trap drive–induced a.c.\ magnetic field is relaxed compared to other species. This highlights HCI as highly promising candidates for next-generation optical clocks with ultra-high accuracy.
The demonstrated method for measuring transverse a.c.\ magnetic fields in multilevel systems is applicable to other singly charged ions with similar level structures.

%This demonstrates that for an optical clock using HCI without hyperfine structure, a measurement of the trap drive induced a.c.\ magnetic field with low uncertainty is not necessary, %unlike for many singly charged-ions-based systems. 
%making HCI ideal candidates for high-precision optical clocks.

\begin{acknowledgments}
    All co-authors declare no conflicts of interest. 
The project was supported by the Physikalisch-Technische Bundesanstalt, the Max-Planck Society, the Max-Planck–Riken–PTB–Center for Time, Constants and Fundamental Symmetries, and the Deutsche Forschungsgemeinschaft (DFG, German Research Foundation) through SCHM2678/5-2, the collaborative research centres SFB 1225 ISOQUANT and SFB 1227 DQ-mat, and under Germany’s Excellence Strategy – EXC-2123 QuantumFrontiers – 390837967. The project has received funding from the EMPIR programme co-financed by the Participating States and from the European Union’s Horizon 2020 research and innovation programme (Projects No. 20FUN01 TSCAC and 23FUN03 HIOC). This project has received funding from the European Research Council (ERC) under the European Union’s Horizon 2020 research and innovation programme (grant agreement No 101019987). The isotopes
for the \Ca production used in this research were supplied by the
U.S. Department of Energy Isotope Program, managed
by the Office of Isotope R\&D and Production.
\end{acknowledgments}

\section*{Data Availability}
The data that support the findings of this article are not publicly available. The data are available from the authors upon reasonable request.
\vspace{5mm}

\appendix
\renewcommand{\thefigure}{A\arabic{figure}}
\setcounter{figure}{0}
\section{\label{appendix}Monte-Carlo sampling for $B_z$}

\begin{figure}[htpb]
\centering
\includegraphics{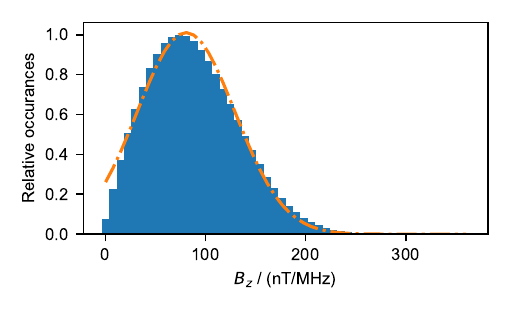}
\caption{Result of Monte-Carlo sampling of $B_z$ as described in the text. The simulation result (blue) is approximated by a Gaussian distribution (orange).}
\label{fig:MC_Bz}
\end{figure}

Solving Eq.~(\ref{eq:Brf_norm}) of the main text for $\langle B_{\mathrm{z, norm}}^2\rangle$ using the measured values of $B_{\mathrm{+, norm}}$, $B_{\mathrm{-, norm}}$, and $\delta\nu_{0\rightarrow0}$ yields nonphysical results with $\langle B_{\mathrm{z, norm}}^2\rangle<0$. However, the experimental uncertainties in the measured values allow for physically valid solutions. 
To estimate the distribution of these possible values, we perform a Monte-Carlo simulation by drawing random samples of $B_{\mathrm{+, norm}}$, $B_{\mathrm{-, norm}}$, and $\delta\nu_{0\rightarrow0}$ from normal distributions centered on their measured values with the corresponding uncertainties. For each sample, we compute $\langle B_{\mathrm{z, norm}}^2\rangle$; unphysical results with $\langle B_{\mathrm{z, norm}}^2\rangle<0$, are discarded when calculating the corresponding $B_{\mathrm{z, norm}}$ values. 
The resulting distribution of $B_z$ is well approximated by a Gaussian distribution, from which we extract its most probable value and uncertainty: $B_{\mathrm{z, norm}}=$ \SI{80\pm49}{\nano\tesla\per\mega\hertz}, as shown in Fig.~\ref{fig:MC_Bz}.

%\section{Appendix B: splitting with different $P_\mathrm{rf}$}
%Setting the magnetic field $B_\mathrm{dc}$ to the resonance, we vary the rf power $P_\mathrm{rf}$. This changes the coupling strength characterized by the on-resonant Rabi frequency $\Omega_\mathrm{ac} \propto \sqrt{P_\mathrm{rf}}$. The signals for different $P_\mathrm{rf}$ are shown in Fig.~\ref{fig:RFDependence}. 
%\begin{figure}[ht]
%    \centering  \includegraphics[width=0.7\columnwidth]{Figures/fig52.pdf}
%    \caption{Experimental excitation probability for changing probe laser detuning $\Delta$ and rf power $P_\mathrm{rf}$ with the trap drive resonant with the Zeeman sublevels.}
%    \label{fig:RFDependence}
%\end{figure}

	%\bibliographystyle{apsrev4-2} % Tell bibtex which bibliography style to use
	%\bibliography{Bibliography.bib} % Tell bibtex which .bib file to use (this one is some example file in TexLive's file tree)
	%apsrev4-2.bst 2019-01-14 (MD) hand-edited version of apsrev4-1.bst
%apsrev4-2.bst 2019-01-14 (MD) hand-edited version of apsrev4-1.bst
%Control: key (0)
%Control: author (72) initials jnrlst
%Control: editor formatted (1) identically to author
%Control: production of article title (-1) disabled
%Control: page (0) single
%Control: year (1) truncated
%Control: production of eprint (0) enabled
%

%Control: key (0)
%Control: author (72) initials jnrlst
%Control: editor formatted (1) identically to author
%Control: production of article title (-1) disabled
%Control: page (0) single
%Control: year (1) truncated
%Control: production of eprint (0) enabled

\end{document}